\documentclass[letterpaper,10pt,conference,twocolumn]{IEEEtran}

\pdfoutput=1
\IEEEoverridecommandlockouts
\usepackage[pdftex]{graphicx}
\DeclareGraphicsExtensions{.pdf,.jpg,.png,.mps}

\usepackage{amsmath,amssymb}  
\usepackage[utf8]{inputenc} 
\usepackage[english]{babel}    

\usepackage{algpseudocode}

\algnewcommand{\LineComment}[1]{\State \(\triangleright\) #1}
\newcommand{\set}[1]{\mathcal{#1}}
\newcommand{\op}[1]{\mathrm{#1}}

\usepackage{amsthm}

\newtheorem{prop}{Proposition}

\theoremstyle{definition}
\newtheorem{example}{Example}
\newtheorem{defi}{Definition}
\newtheorem{problem}{Problem}

\usepackage{float}
\floatstyle{ruled}
\newfloat{algorithm}{thp}{lop}
\floatname{algorithm}{Algorithm}

\usepackage{pgf}

\begin{document}

\title{Coordinating Truck Platooning by Clustering Pairwise Fuel-Optimal Plans}
\author{Sebastian van de Hoef, Karl H. Johansson and Dimos V. Dimarogonas
\thanks{The authors are with the ACCESS Linnaeus Center and the School of Electrical Engineering, KTH Royal Institute of Technology, SE-100 44, Stockholm, Sweden. {\tt \{shvdh, kallej, dimos\}@kth.se}. This work was supported by the COMPANION EU project, the Knut and Alice Wallenberg Foundation, and the Swedish Research Council.
}
}

\maketitle                    

\begin{abstract}
 We consider the fuel-optimal coordination of trucks into platoons. Truck platooning is a promising technology that enables trucks to save significant amounts of fuel by driving close together and thus reducing air drag. We study how fuel-optimal speed profiles for platooning can be computed. A first-order fuel model is considered and pairwise optimal plans are derived. We formulate an optimization problem that combines these pairwise plans into an overall plan for a large number of trucks. The problem resembles a medoids clustering problem. We propose an approximation algorithm similar to the partitioning around medoids algorithm and discuss its convergence. The method is evaluated with Monte Carlo simulations. We demonstrate that the proposed algorithm can compute a plan for thousands of trucks and that significant fuel savings can be achieved.
\end{abstract}

\section{Introduction}

Platooning is a common element in intelligent transportation systems. It refers to a group of vehicles forming a road train without any physical coupling between the vehicles. A short inter-vehicle distance is maintained by automatic control and vehicle-to-vehicle communication. 
Platooning is beneficial in various ways. Due to the small inter-vehicle gaps the total road throughput can be increased \cite{path_overview_conference}. It can also help facilitate the driverless operation of the trailing vehicles. 
In this paper, our main motivation to study platooning is the potential to reduce the fuel consumption of the follower vehicles in the platoon.  
Similar to what racing cyclists exploit, the follower vehicles experience a reduction in air drag which translates into reduced fuel consumption \cite{Bonnet2000}. 

Using platooning to reduce fuel consumption leads to a challenging coordination problem. This differs from a classical setting where platoons are primarily a measure to increase road throughput and where a majority of the vehicles can platoon. Consider two trucks that travel between the same two regions but from different locations within the regions and at approximately the same time. Then the trucks can adjust their speeds slightly at the beginning of their journeys, form a platoon at the start of the common part of their route and thus save fuel during most of their journeys. The catch is that this might involve one of the trucks having to drive faster before the two merge, which increases air-drag and consequently fuel consumption during that phase. One truck might also slow down to let the other truck catch up but then travel at an increased speed later on to arrive at its destination on time. If many trucks are involved, it is not straightforward to compute a plan for all trucks that would be globally fuel-optimal. This is the problem we address in this paper.

Variations of this problem have been considered in literature. In \cite{jeff_complexity} the authors formulate a mixed integer linear programming problem, without considering the speed dependency of fuel consumption, and prove that the problem is NP-hard. In \cite{kuoyun_catchup} the authors consider a simple catch-up coordination scheme and evaluate it on real fleet data. In \cite{jeff_kuo_yun_distributed_controller} local controllers for coordinating the formation of platoons are proposed. In \cite{datamining_platooning} the authors use data-mining to identify economic platoons based on different criteria. They allow that trucks wait for other trucks to form the platoon. In \cite{ACCpaper} the problem is formulated as an optimization problem. The proposed method is, however, not applicable to large numbers of trucks due to the combinatorial complexity.

Clustering is present in a variety of different contexts. A large body of research focuses on clustering methods as an analysis tool to structure, understand, and classify large data sets \cite{clustering_book1, clustering_overview_paper}. Examples include the clustering of graphs in the scope of community detection \cite{blondelcommunity, community_detection_survey, community_detection_survey2}. Our clustering algorithm is inspired by partitioning around medoids \cite{pam_book}.
A closely related area to our subproblem of choosing coordination leaders is leader election, where a group of agents has to jointly determine a leader \cite{bully_alg, electing_good_leaders}. When we interpret pairwise fuel savings as preferences for trucks being coordination leaders, we can see our algorithm as local leader election.
In one variant of the proposed clustering algorithm we consider that leader and follower share their benefit from platooning according to a fixed ratio. This setting is related to coalitional game theory \cite{coalitional_game_theory}, which is a widely used framework in communication theory.

We derive a powerful, efficient, and scalable method to coordinate platooning of a large number of trucks in a fuel-optimal way. The main contribution is that the coordination problem is formulated as a clustering problem based on pairwise fuel-optimal speed profiles. We derive pairwise fuel-optimal plans based on a first-order fuel model and propose a clustering algorithm. We discuss the convergence of the clustering algorithm and demonstrate the entire method using Monte Carlo simulations. 

The paper is organized as follows. We start in Section~\ref{sec:modeling} by modeling a single truck as a hybrid system, where the discrete jumps model transitions between road segments. In Section~\ref{sec:optimal_speed_adaption} we develop a fuel-optimal plan for two trucks, where the leader keeps a constant speed and the the follower adapts its speed to merge with the leader. In Section~\ref{sec:graph_based_clustering} we build on these pairwise plans to construct a similarity graph and formulate the problem to select good leaders and match followers to each of them. In Section~\ref{sec:approxmation_algs} we propose a clustering algorithm to find approximate solutions to the problem and analyze its convergence. In Section~\ref{sec:simulations} the method is demonstrated by Monte Carlo simulations.

\section{Modeling}
\label{sec:modeling}

In this section we model a single truck as a hybrid system, define platooning, and introduce a fuel consumption model.

We consider a highway network in which a set of geographic locations are connected by one-way road segments.
We model the road network as a weighted, directed graph $\set{G} = (\set{N},\set{E},W)$, where the nodes $\set{N}$ correspond to geographic locations. An edge $(i,j) \in \set{E} \subset \set{N} \times \set{N}$ models a road segment from the location of node $i$ to the location of node $j$. The weight $W(e) > 0$ of edge $e \in \set{E}$ models the length of the road segment. The state of a truck $(x,\ell)$ comprises the edge $\ell \in \set{E}$ that corresponds to the truck's current road segment and the distance $x \in [0,W(\ell)]$ from the beginning of that road segment.

\begin{figure}
\begin{center}
 \def\svgwidth{\columnwidth}
 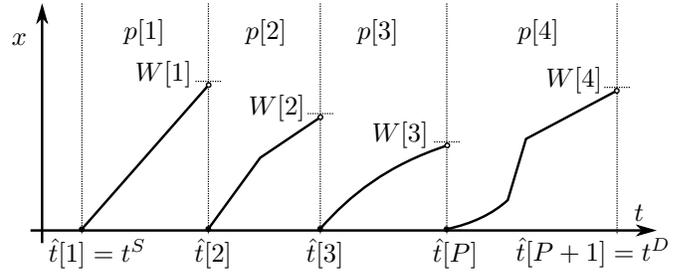
 \caption{A trajectory of the hybrid truck model}
 \label{fig:hybrid_solution}
\end{center}
\end{figure}

\begin{defi}[Hybrid Truck Model]\label{def:hybrid_truck_model}
A truck is modeled as a hybrid system with flow map \eqref{eq:flow_map}, flow set $\{x \leq W(\ell(t))\}$, jump map \eqref{eq:jump_map}, and jump set $\{x \geq W(\ell(t))\}$:
\begin{align}
 &\dot{x}(t) = v(t), && \label{eq:flow_map}\\
 &x^+ = 0, &&\ell^+ \in \{(i,j) \in \set{E}: (\cdot,i) = \ell\}, \label{eq:jump_map}
\end{align}
where the speed $v(t) \in [v_{\min},v_{\max}]$ and the next edge $\ell^+$ are control inputs. We assume $v_{\max} \geq v_{\min} > 0$, i.e., trucks cannot stop or travel backwards.
\end{defi}
As long as the system state $(x,\ell)$ is in the flow set, $x$ evolves continuously according to \eqref{eq:flow_map}. When the state reaches the jump set, i.e., when the truck reaches the end of the segment, the state is reset according to the jump map \eqref{eq:jump_map}. Figure \ref{fig:hybrid_solution} illustrates the trajectory of such a system. For theoretical considerations of such models see \cite{hybridbook}.

A truck has a transport assignment to travel through the road network from a start node $n^\op{S} \in \set{N}$ at time $t^\op{S}$ to a destination node $n^\op{D} \in \set{N}$ where it must arrive at time $t^\op{D}$. We assume that there is a unique shortest path $\bar{p} = \bar{p}[1],\bar{p}[2],\dots,\bar{p}[P]$, $\bar{p}[i] \in \set{E}$ for $i \in \{1,\dots,P\}$ from $n^\op{S}$ to $n^\op{D}$ with respect to $W$ and we let the truck travel along that path: $\ell^+(\bar{p}[i]) = \bar{p}[i+1]$. With this notation we can define the trajectory of a truck (compare to Figure \ref{fig:hybrid_solution}). 
\begin{defi}[Trajectory]\label{def:trajectory}
 We call $(x(t),l(t)): [t^\op{S},t^\op{D}) \rightarrow \mathbb{R} \times \set{E}$ a trajectory of a truck if there exists a finite series of jump times $\hat{t}[1] < \hat{t}[2] < \dots < \hat{t}[P] < \hat{t}[P+1]$ with $\hat{t}[1] = t^\op{S}$, $\hat{t}[P+1] = t^\op{D}$ and for $i \in \{1,\dots,P\}$: $x(\hat{t}[i]) = 0$, $\lim\limits_{t \nearrow \hat{t}[i+1]}x(t) = W(\bar{p}[i])$ and for $t \in [\hat{t}[i],\hat{t}[i+1])$ it holds that $x(t) \leq W(\bar{p}[i])$, $\ell(t) = \bar{p}[i]$, and $\dot{x}(t) = v(t)$. 
\end{defi}

In the following we introduce a system composed of $K$ trucks. We introduce for each truck $k \in \{1,\dots,K\}$ its state $(x_k,\ell_k)$.  Truck $k$ starts at node $n_k^\op{S}$ at time $t_k^\op{S}$ and must arrive at node $n_k^\op{D}$ at time $t_k^\op{D}$. Its shortest path is denoted $\bar{p}_k$ and has length $P_k$. Its speed is denoted $v_k$. We have the start position $(x_k^\op{S},\ell_k^\op{S}) = (0,\bar{p}_k[1])$ and the destination position $(x_k^\op{D},\ell_k^\op{D}) = (W(\bar{p}_k[P_k]),\bar{p}_k[P_k])$. 
 
\subsection*{Platooning}

Next, we formalize platooning. For the purpose of coordination we neglect the small inter-vehicle distance and let trucks that platoon with each other assume the same position.

Let $\set{P} \subset \{1,\dots,K\}$ be a set of trucks. Let $(x_k(t),\ell_k(t))$ be the trajectory of truck $k \in \set{P}$. If trucks $\set{P}$ platoon from time $t^\op{M}$ to $t^\op{Sp}$, then $(x_i(t),\ell_i(t)) = (x_j(t),\ell_j(t))$ for all pairs $(i,j) \in \set{P}\times\set{P}$ and all $t \in [t^\op{M},t^\op{Sp}]$. 
We assign exactly one truck in the platoon the role of a \textit{platoon leader} and the other trucks the role of \textit{platoon followers}. Only the fuel consumption of platoon followers is reduced.

We model the fuel consumption per distance traveled as first-order polynomials of the speed.
We believe that this type of model approximates the real fuel consumption in the interval $[v_{\min}, v_{\max}]$ well enough for our purpose.
When a truck is not in a platoon or is a platoon leader, its fuel consumption is $f_0(v) = F^1 v + F^0$ where $v$ is the speed and $F^1, F^0 \in \mathbb{R}$ are constants. When a truck is a platoon follower the fuel consumption is modeled by $f_p(v) = F_p^1 v + F_p^0$ where $F_p^1$, $F_p^0$ are other constants. Typically, $F^1 > F_p^1 > 0$ and $f_0(v) > f_p(v)$ for $v \in [v_{\min}, v_{\max}]$.
The total fuel consumption of truck $k$ from start to destination is
\begin{equation}
 f_{\op{c},k} = \sum\limits_{i = 1}^{P_k} \int\limits_{0}^{W(\ell_k[i])} f_k\big(v_k(t_k(x,\ell_k[i])),t_k(x,\ell_k[i])\big) \op{d}x, \label{eq:fuel_consumption}
\end{equation}
where $f_k(v,t) = f_p(v)$, if the truck is a platoon follower at time $t$, and $f_k(v,t) = f_0(v)$ otherwise. The function $t_k(x,\ell_k[i])$ is the time instant at which truck $k$ is at position $(x,\ell_k[i])$.

\section{Optimal Speed Adaptation for Platooning}
\label{sec:optimal_speed_adaption}

\begin{figure}
\begin{center}
 \def\svgwidth{\columnwidth}
 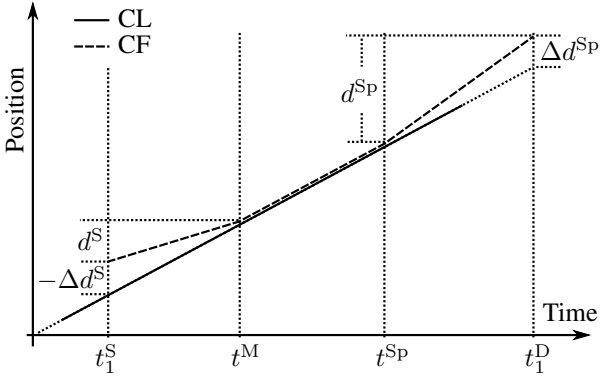
 \caption{Speed profiles of the CL and the CF.}
 \label{fig:speed_profile}
\end{center}
\end{figure}

In this section we derive a pairwise fuel-optimal speed plan. We consider a pair of trucks: a \textit{coordination leader} (CL) and a \textit{coordination follower} (CF). These concepts differ from the concepts of a platoon leader and platoon follower that were introduced in Section~\ref{sec:modeling}. The CL keeps a constant speed while the CF selects at the beginning of its journey a speed that allows it to merge into a platoon with the CL. Then the two platoon until they split up, followed by that the CF selects a speed so that it arrives at its pre-specified arrival time at its destination. Figure~\ref{fig:speed_profile} illustrates these three phases. We first derive the optimal speed during the first and the last phase neglecting that trucks can only platoon on common road segments. We add this constraint in the second part of the section. In Section~\ref{sec:graph_based_clustering}, we build on such pairwise plans and obtain an overall plan for all $K$ trucks. First, we define the default speed profile, i.e., a constant speed.
\begin{defi}[Default Speed Profile]
The default speed profile for a truck $k$ is the constant speed
\begin{align}
 v_{k}(t) = \frac{ \sum\limits_{i=1}^{P_k} W(\bar{p}_k[i]) }{ t_{k}^\op{D} - t_{k}^\op{S}}, \label{eq:leader_velocity}
\end{align}
for $t \in [t_k^\op{S},t_k^\op{D})$.
The resulting fuel consumption is $\bar{f}_{\op{c},k}$. 
\end{defi}

The following proposition gives the optimal rendezvous speed for a CF when there are no restrictions as to where the CL and the CF can meet.
\begin{prop}\label{prop:unconstrained_vs}
  Assume the following.
  The speed of truck $0$ is constant $\dot{x}_0 = v_0$ with $v_0 \in \mathbb{R}$, $v_0 > 0$. The state of truck $0$ at time $t^\op{S}$ is $(x_0(t^\op{S}),\ell)$. 
  The state of truck $1$ at time $t^\op{S}$ is $(x_1(t^\op{S}),\ell)$. 
  Truck $1$ platoons with truck $0$ between time $t^\op{M}$ and $t^\op{Sp}$ with $t^\op{Sp} > t^\op{M}$. Truck $1$ has constant speed $\dot{x}_1 = v_\op{s}$ for time $t^\op{S}$ to $t^\op{M}$ and $\dot{x}_1 = v_0$ from time $t^\op{M}$ to $t^\op{Sp}$. The rendezvous speed $v_\op{s}$ is constrained to the interval $[v_{\min},v_{\max}]$. The road segment corresponding to $\ell$ is sufficiently long. 
  
  Then the rendezvous speed $v_\op{s}^*$ that minimizes fuel consumption from time $t^\op{S}$ to $t^\op{Sp}$ is given by
  \begin{equation}
  \begin{split}
  &v_\op{s}^* =\\
  &\left\{
  \begin{array}{ll}
    \max \left(v_{0}\left(1 - \sqrt{1 - \frac{F_p^1}{F^1} + \frac{\Delta F^0}{F^1 v_{0}}}\right), v_{\min}\right) &\text{if } \Delta d < 0\\
    \min \left(v_{0}\left(1 + \sqrt{1 - \frac{F_p^1}{F^1} + \frac{\Delta F^0}{F^1 v_{0}}}\right), v_{\max}\right) &\text{if } \Delta d > 0\\
    v_{0} &\text{if } \Delta d = 0,
  \end{array}
  \right.
  \end{split}
  \end{equation}
  where $\Delta d = x_0(t^\op{S}) - x_1(t^\op{S})$ and $\Delta F^0 = F^0 - F_p^0$.
\end{prop}
The proof of Proposition \ref{prop:unconstrained_vs} can be found in the Appendix.

The optimal speed after splitting up for the CF to meet its deadline is also the appropriate $v_\op{s}^*$ depending on whether the CF has to speed up or slow down in order to arrive at the specified time. 

In the remainder of this section, we discuss how Proposition~\ref{prop:unconstrained_vs} can be used to compute an optimal speed profile when the CL and the CF travel on different but intersecting paths. Figure~\ref{fig:speed_profile} illustrates the discussion. 

To this end, we define the distance $d_p$ between two states $(x_1,\ell_1)$, $(x_2,\ell_2)$ with respect to a path $p$. 
\begin{defi}[Distance]
Let $i_1$ be such that $p[i_1] = \ell_1$ and $i_2$ be such that $p[i_2] = \ell_2$ and assume $i_2 \geq i_1$. Then,
\begin{equation}
  d_p\big((x_1,\ell_1),(x_2,\ell_2)\big) =
   \left|x_2 - x_1 + \sum\limits_{i = i_1}^{i_2 -1} W(p[i])\right|
\end{equation}
\end{defi}
Consider a CL with index $0$ and a CF with index $1$.
Two trucks can platoon only on the road segments corresponding to common edges of their paths. Because their paths are shortest paths it can be shown that the shared edges between two paths form a path as well (Lemma 1 in \cite{ACCpaper}), i.e., two paths meet and split up at most once. Trucks $0$,$1$ start at $(x_0^\op{S},\ell_0^\op{S})$, $(x_1^\op{S},\ell_1^\op{S})$ at time $t_0^\op{S}$, $t_1^\op{S}$ and arrive at $(x_0^\op{D},\ell_0^\op{D})$, $(x_1^\op{D},\ell_1^\op{D})$ at time $t_0^\op{D}$, $t_1^\op{D}$, respectively.
We denote the position at which the CL and the CF start platooning at time $t^\op{M}$ as $(x^\op{M},\ell^\op{M})$ and where they split at time $t^\op{Sp}$ as $(x^\op{Sp},\ell^\op{Sp})$. These meeting points have to lie on the trajectory of the CL with constant speed $v_{0}$:
\begin{align*}
 d_{\bar{p}_0}\big((x_0^\op{S},\ell_0^\op{S}),(x^\op{M},\ell^\op{M})\big) &= v_{0} (t^\op{M} - t_0^\op{S})\\
 d_{\bar{p}_0}\big((x_0^\op{S},\ell_0^\op{S}),(x^\op{Sp},\ell^\op{Sp})\big) &= v_{0} (t^\op{Sp} - t_{0}^\op{S}).
\end{align*}

When platooning with the CL the planned trajectory of the CF consists of three phases: from start to the meeting point with speed $v^\op{S}$, from meeting point to the split point platooning as platoon follower of $0$ with speed $v_{0}$, and from the split point to the destination with speed $v^\op{Sp}$. We define $d^\op{S} = d_{\bar{p}_1}\big((x_1^\op{S},\ell_1^\op{S}),(x^\op{M},\ell^\op{M})\big)$ and $d^\op{Sp} = d_{\bar{p}_1}\big((x^\op{Sp},\ell^\op{Sp}),(x_1^\op{D},\ell_1^\op{D})\big)$. We have the relations
\begin{align*}
 d^\op{S} = v^\op{S}(t^\op{M} - t_1^\op{S}),\;\;\;
 d^\op{Sp} = v^\op{Sp}(t_1^\op{D} - t^\op{Sp}).
\end{align*}
We define the virtual position difference at the start/end of the CF's trajectory as
\begin{equation}
\begin{split}
 \Delta d^\op{S} &= d^\op{S} - (t^\op{M} - t_{1}^\op{S})v_{0}\\
 \Delta d^\op{Sp} &= d^\op{Sp} - (t_{1}^\op{D} - t^\op{Sp})v_{0}, \label{eq:delta_d_general}
\end{split}
\end{equation}
which is equivalent to $\Delta d$ in Proposition~\ref{prop:unconstrained_vs}. If $\Delta d^\op{S} > 0$ then $v^\op{S} > v_{0}$, if $\Delta d^\op{S} < 0$ then $v^\op{S} < v_{0}$, if $\Delta d^\op{Sp} > 0$ then $v^\op{Sp} > v_{0}$, and if $\Delta d^\op{Sp} < 0$ then $v^\op{Sp} < v_{0}$. Then, we can compute according to \eqref{eq:v_star_def} the appropriate $v^*$ for the first and the last phase. 

This derivation has ignored so far that the first possible point to merge is when the CL's and the CF's paths meet. If $v^*$ leads to a distance from $(x_1^\op{S},\ell_1^\op{S})$ to the merge point that is too small, then the CL selects a speed that lets the CL and CF merge at the position where the two paths meet, denoted here $(0,\ell^F)$. 
This speed is $v^\op{S} = d_{\bar{p}_1}\big((x_1^\op{S},\ell_{1}^\op{S}),(\ell^F,0)\big)/(t^\op{M} - t_{1}^\op{S})$. The corresponding case might occur at split up, so that $v_{1}^\op{Sp} = d_{\bar{p}_1}\big((W(\ell^L),\ell^L),(x_{1}^\op{D},\ell_{1}^\op{D})\big)/(t_{1}^\op{D} - t^\op{Sp})$, where $(W(\ell^L),\ell^L)$ is the position where the CL's and the CF's paths split up.

The first check to test if platooning is possible and beneficial is, whether the calculated merge point lies before the split point or not, i.e., whether
$
 d^\op{S} + d^\op{Sp} < d_{\bar{p}_1}\big((x_{1}^\op{S},\ell_{1}^\op{S}),(x_{1}^\op{D},\ell_{1}^\op{D})\big).
$
If this condition is fulfilled, we can calculate the fuel cost for the CF with the speed profile that is adapted for platooning with the CL as follows
\begin{equation}
\begin{split}
 f_{p} &= d^\op{S} f_0(v^\op{S}) + d^\op{Sp} f_0(v^\op{Sp})\\
 &+ \big(d_{\bar{p}_1}\big((x_1^\op{S},\ell_1^\op{S}),(x_1^\op{D},\ell_1^\op{D})\big) - d^\op{S} - d^\op{Sp}\big) f_p(v_{0}). \label{eq:adapted_fuel_consumption}
\end{split}
\end{equation}

We summarize the results of the Section in definition of the adapted speed profile.
\begin{defi}[Adapted Speed Profile]
 The adapted speed profile of a CF with index $1$ to a CL with index $0$ consists of three phases with constant speed: $v^\op{S}$ from $t_1^\op{S}$ to $t^\op{M}$, then $v_0$ from $t^\op{M}$ to $t^\op{Sp}$, and finally $v^\op{Sp}$ from $t^\op{Sp}$ to $t_1^\op{D}$, where the CF is platoon follower of the CL from time $t^\op{M}$ to $t^\op{Sp}$. The resulting fuel consumption is $f_{p}$ as in \eqref{eq:adapted_fuel_consumption}. Jumps in the speed at time $t^\op{M}$ and $t^\op{Sp}$ are continuously approximated.
\end{defi}
 
\section{Coordination Leader Selection}
\label{sec:graph_based_clustering}

In this section we combine default speed profiles and adapted speed profiles from Section~\ref{sec:optimal_speed_adaption} in order to minimize the total fuel consumption $\sum\limits_{k = 1}^K f_{\op{c},k}$. We select a number of CLs that stick to their default speed profile. All other trucks are assigned to their best CL if there is one. The best CL for a CF is the one that yields the largest fuel saving for the CF. Since the speed profile of the CL is not affected by the CF, we can assign several CFs to a CL independently. Note that the method presented in this section can accommodate default and adapted speed profiles different from the ones presented in Section~\ref{sec:optimal_speed_adaption}. Like this factors such as traffic, different fuel models, soft constraints on the arrival time, etc. can be considered.

Based on the results from Section~\ref{sec:optimal_speed_adaption}, we calculate the potential fuel savings gained from platooning for all pairs of trucks, where one takes the role of the CL and the other the role of the CF. We collect the results in a weighted directed graph, which we call the coordination graph. Recall that the total fuel consumption of a truck $i$ with the default speed profile is denoted $\bar{f}_{\op{c},k}$. We denote the fuel consumption of a CF $i$ with CL $j$ as $f_{\op{p},(i,j)}$. If $i$ and $j$ platooning with $i$ being the CF of $j$ is not possible, then $f_{\op{p},(i,j)} = \bar{f}_{\op{c},k}$.
\begin{defi}[Coordination Graph]
 The coordination graph is a weighted directed graph $\set{G}_\op{c} = (\set{N}_\op{c},\set{E}_\op{c},W_\op{c})$. $\set{N}_\op{c}$ is a set of $K$ nodes, where each node represents a truck. $\set{E}_\op{c} \subseteq \set{N}_\op{c} \times \set{N}_\op{c}$ is a set of edges, and $W_\op{c}: \set{E}_\op{c} \rightarrow \mathbb{R}^+$ are non-negative edge weights. There is an edge $(i,j)$, if $i$ saves fuel when it selects $j$ as CL, i.e., $\set{E}_\op{c} =  \{(i,j) \in \set{N}_\op{c} \times \set{N}_\op{c}: f_{\op{p},(i,j)} < \bar{f}_{\op{c},i}, i \neq j\}$, and $W_\op{c}\big((i,j)\big) = \bar{f}_{\op{c},i} - f_{\op{p},(i,j)}$. If platooning is not possible or beneficial, there is no edge. 
\end{defi}

We introduce the set of in-neighbors of a node $n$ as $\set{N}_{\op{i}}(n) = \{i \in \set{N}_\op{c}: \exists (i,n) \in \set{E}_\op{c}\}$ and the set of out-neighbors of a node $n$ as $\set{N}_{\op{o}}(n) = \{i \in \set{N}_\op{c}: \exists (n,i) \in \set{E}_\op{c}\}$. We define that the maximum over an empty set is zero, i.e., $\max\limits_{i \in \emptyset}(\cdot) = 0$. 

Next, we formulate the problem of finding a fuel optimal set of CLs $\set{N}_\op{l}$.
\begin{problem}\label{prob:clustering}
Find a subset $\set{N}_\op{l} \subset \set{N}_\op{c}$ of nodes that maximizes $f_\op{ce}(\set{N}_\op{l})$ where
\begin{equation}
 f_\op{ce}(\set{N}_\op{l}) = \sum\limits_{i \in \set{N}_\op{c} \setminus \set{N}_\op{l}} \max_{j \in \set{N}_{\op{o}}(i) \cap \set{N}_\op{l}}  W_\op{c}(i,j).  \label{eq:clustering_problem}
\end{equation}
\end{problem}
If $(i,j) \in \set{E}_\op{c}$ with $i \in \set{N}_\op{c} \setminus \set{N}_\op{l}$ and $j = \arg \max\limits_{j \in \set{N}_\op{o}(i) \cap \set{N}_\op{l}}(W_\op{c}(i,j))$, we say that $i$ is the CF of $j$ and $j$ is the CL of $i$. If $j$ has no out-neighbor in $\set{N}_\op{l}$, then $\max_{j \in (\set{N}_{\op{o}}(i) \cap \set{N}_\op{l})} ( W_\op{c}(i,j) ) = \max_{j \in \emptyset} ( W_\op{c}(i,j) ) = 0$. 

This problem is similar to k-medoids clustering \cite{pam_book}. Medoids clustering means selecting a, typically fixed, number of cluster centers from a set of data points. The remaining data points are assigned to the closest cluster center. The objective is to find cluster centers in such a way that the sum of the distances of all nodes to the closest cluster center is minimal. In our problem we can identify the CLs with the medoids. 

Albeit similar to k-medoids clustering, there are some significant differences. We maximize similarity within the cluster as opposed to minimizing the distance. Similarity from node $i$ to node $j$ is here how much fuel is saved if node $i$ selects node $j$ as a CL. This implies that the similarity between two nodes is not symmetrical, non-negative, and can be zero.  The number of clusters is not fixed but part of the optimization and not every node has to be assigned to a cluster. It is easy to see, that a solution with zero or $K$ clusters is not optimal if $\set{E}_\op{c} \neq \emptyset$. This is different from an application where the sum of the distances to the cluster centers is minimized. If every node is a cluster center, this trivially corresponds to the smallest possible objective value. Thus, it is not trivial to find (approximate) solutions to the problem.

\section{Leader Selection Clustering Algorithm}
\label{sec:approxmation_algs}

In this section, we describe four variants of an approximation algorithm (Algorithm~\ref{alg:general_alg}) to Problem~\ref{prob:clustering} inspired by partitioning around medoids (PAM) \cite{pam_book}. 
Algorithm~\ref{alg:general_alg} works according to the following scheme. At the beginning $\set{N}_\op{l}$ is empty, i.e., there are no CLs. In each round a node $n \in \set{N}_\op{c}$ is selected for which it is beneficial to be added to $\set{N}_\op{l}$ or removed from $\set{N}_\op{l}$, and $\set{N}_\op{l}$ is updated accordingly. This is repeated until no further improvement is possible. We introduce a function $\Delta u(n,\set{N}_\op{l})$ that measures how much is gained from switching whether $n$ belongs to $\set{N}_\op{l}$ or not.

\begin{algorithm}
\begin{algorithmic}
\Require $\set{G}_\op{c}$
\Ensure $\set{N}_\op{l}$

\State $\set{N}_\op{l} \leftarrow \emptyset$
\While{$\{\bar{n} \in \set{N}_\op{c}: \Delta u(\bar{n},\set{N}_\op{l}) > 0\} \neq \emptyset$}
  \State Select $n \in \{\bar{n} \in \set{N}_\op{c}: \Delta u(\bar{n},\set{N}_\op{l}) > 0\}$
    \If {$n \in \set{N}_\op{l}$}
      \State $\set{N}_\op{l} \leftarrow \set{N}_\op{l} \setminus \{n\}$
    \Else
      \State $\set{N}_\op{l} \leftarrow \set{N}_\op{l} \cup \{n\}$
    \EndIf
\EndWhile
\end{algorithmic}
\caption{}\label{alg:general_alg}
\end{algorithm}

We consider two methods to select $n$ and two choices of $\Delta u(n,\set{N}_\op{l})$.
The first method for selecting $n$ is to select it in a greedy manner according to $n = \arg \max\limits_{\bar{n} \in \set{N}_\op{c}} \Delta u(\bar{n},\set{N}_\op{l})$. This corresponds to what is done in the ``build phase'' of PAM. The second method is to choose $n$ randomly with equal probability from the set $\{\bar{n} \in \set{N}_\op{c}: \Delta u(\bar{n},\set{N}_\op{l}) > 0\}$. The two different choices for $\Delta u(n,\set{N}_\op{l})$ are elaborated in the following Sections~\ref{sec:central_objective}~and~\ref{sec:individual_objective}.

\subsection{Total Gain}
\label{sec:central_objective}

The most obvious choice for $\Delta u(n,\set{N}_\op{l})$ is to consider how much $f_\op{ce}$ as defined in \eqref{eq:clustering_problem} changes. This can be calculated locally, which means only considering one-hop and two-hop neighbors of $n$. We get, if $n \notin \set{N}_\op{l}$,
\begin{align*}
 &f_\op{ce}(\set{N}_\op{l} \cup \{n\}) - f_\op{ce}(\set{N}_\op{l})=\\
 &\sum\limits_{i \in \set{N}_{\op{i}}(n)\setminus \set{N}_\op{l}} \left( 
 \max\limits_{j \in \set{N}_{\op{o}}(i) \cap (\set{N}_\op{l} \cup \{n\})} W_\op{c}(i,j)
 - \max\limits_{j \in \set{N}_{\op{o}}(i) \cap \set{N}_\op{l}} W_\op{c}(i,j)
 \right)\\
 &- \max\limits_{i \in \set{N}_{\op{o}}(n) \cap \set{N}_\op{l}} W_\op{c}(n,i).
\end{align*}
The sum over $i$ covers nodes that can select $n$ as their new CL. The last summand accounts for $n$ possibly not being a CF any longer.
Similarly, if $n \in \set{N}_\op{l}$,
\begin{align*}
 &f_\op{ce}(\set{N}_\op{l} \setminus \{n\}) - f_\op{ce}(\set{N}_\op{l}) =\\
 &\sum\limits_{i \in \set{N}_{\op{i}}(n)\setminus \set{N}_\op{l}} \left( 
 \max\limits_{j \in \set{N}_{\op{o}}(i) \cap (\set{N}_\op{l} \setminus \{n\})} W_\op{c}(i,j)
 - \max\limits_{j \in \set{N}_{\op{o}}(i) \cap \set{N}_\op{l}} W_\op{c}(i,j)
 \right)\\
 &+ \max\limits_{i \in \set{N}_{\op{o}}(n) \cap (\set{N}_\op{l} \setminus \{n\})} W_\op{c}(n,i). 
\end{align*}
The sum over $i$ covers nodes that can have $n$ as their CL before the change. The last summand accounts for $n$ possibly becoming a CF. Finally we get
\begin{align}
 \Delta u(n,\set{N}_\op{l}) = \left\{ \begin{array}{ll}
                   f_\op{ce}(\set{N}_\op{l} \setminus \{n\}) - f_\op{ce}(\set{N}_\op{l}) & \text{ if } n \in \set{N}_\op{l}\\
                   f_\op{ce}(\set{N}_\op{l} \cup \{n\}) - f_\op{ce}(\set{N}_\op{l}) & \text{ otherwise }.
                  \end{array} \right.\label{eq:delta_u_c}
\end{align}

\subsection{Pairwise Gain}
\label{sec:individual_objective}

Instead of considering the global objective $f_\op{ce}$, we can consider that the pairwise fuel savings from platooning are divided between CF and CL according to a fixed ratio. We think of this as a simple compensation scheme for trucks from different operators platooning with each other. The trucks acting as platoon leaders need to have an incentive since they do not get any fuel savings from platooning. If $i$ is the CF of $j$, then $j$ gets the utility $\rho_\op{l} W(i,j)$ with $\rho_\op{l} \in (0, 1) \subset \mathbb{R}$ and $i$ gets the utility $\rho_\op{f} W(i,j)$ with $\rho_\op{f} = 1-\rho_\op{l}$ from this CF. The total utility of the CL is the sum of the utilities from all its CFs. The gain $\Delta u(\bar{n},\set{N}_\op{l})$ for a node $\bar{n}$ is the change of its utility from switching whether $n$ belongs to $\set{N}_\op{l}$ or not. 

The utility of a node $i \notin \set{N}_\op{l}$ is 
\begin{align*}
 u_\op{f}(i,\set{N}_\op{l}) = 
 \rho_\op{f} \max\limits_{j \in \set{N}_\op{o}(i) \cap \set{N}_\op{l}}W_\op{c}(i,j).
\end{align*}
With a set of CLs $\set{N}_\op{l}$ given, the CL of a CF, as introduced in Section~\ref{sec:graph_based_clustering}, is the truck in $\set{N}_\op{l}$ that maximizes the utility of a CF. The utility $u_\op{l}(j,\set{N}_\op{l})$ of a CL $j$ is the sum of the utilities due to its CFs $\set{N}_\op{f}(j) = \{ i \in (\set{N}_\op{c}\setminus \set{N}_\op{l}):j = \arg \max_{k \in \set{N}_\op{0}(i) \cap \set{N}_\op{l}}W(i,k)\}$
\begin{align*}
 u_\op{l}(j,\set{N}_\op{l}) = \sum\limits_{i \in \set{N}_\op{f}(j) } \rho_\op{l} W(i,j) = \sum\limits_{i \in \set{N}_\op{f}(j) } \frac{\rho_\op{l}}{\rho_\op{f}} u_\op{f}(i,\set{N}_\op{l}).
\end{align*}
If $n \notin \set{N}_\op{l}$, then the gain by becoming a CL, i.e., adding $n$ to $\set{N}_\op{l}$, is
$u_\op{l}(n,\set{N}_\op{l} \cup \{n\}) - u_\op{f}(n,\set{N}_\op{l})$.
If $n \in \set{N}_\op{l}$, then the gain by becoming a CF, i.e., removing $n$ from $\set{N}_\op{l}$, is
$u_\op{f}(n,\set{N}_\op{l} \setminus \{n\}) - u_\op{l}(n,\set{N}_\op{l})$. The larger $\rho_\op{l}$ the larger the incentive to be a CL.
Finally we have
\begin{align}
 \Delta u(n,\set{N}_\op{l}) = \left\{ \begin{array}{ll}
                   u_\op{f}(n,\set{N}_\op{l} \setminus \{n\}) - u_\op{l}(n,\set{N}_\op{l}) & \text{ if } n \in \set{N}_\op{l}\\
                   u_\op{l}(n,\set{N}_\op{l} \cup \{n\}) - u_\op{f}(n,\set{N}_\op{l}) & \text{ otherwise.}
                  \end{array} \right.\label{eq:delta_u_g}
\end{align}
The gain $\Delta u(n,\set{N}_\op{l})$ can be calculated locally with the knowledge of: the edges of $n$, the utilities of the in-neighbors of $n$, and which neighbors belong to $\set{N}_\op{l}$.
\subsection{Convergence}
In this section we analyze the convergence of Algorithm~\ref{alg:general_alg}. Algorithm~\ref{alg:general_alg} converges when it reaches a state where $\{\bar{n} \in \set{N}_\op{c}: \Delta u(\bar{n},\set{N}_\op{l}) > 0\}$ is empty. We call such a state \textit{equilibrium}.
We have the following result for the central objective.
\begin{prop}
 Algorithm~\ref{alg:general_alg} converges to an equilibrium with $\Delta u$ according to \eqref{eq:delta_u_c} if in each iteration $n$ is selected as $n = \arg \max\limits_{\bar{n} \in \set{N}_\op{c}} \Delta u(\bar{n},\set{N}_\op{l})$. If $n$ is selected randomly with equal probability from $\{\bar{n} \in \set{N}_\op{c}: \Delta u(\bar{n},\set{N}_\op{l}) > 0\}$, Algorithm~\ref{alg:general_alg} converges with probability 1.
\end{prop}
\begin{proof}
 $f_\op{ce}(\set{N}_\op{l})$ increases in each iteration when $n$ with the largest $\Delta u(n,\set{N}_\op{l})$ is selected, but since $\set{N}_\op{l} \subset \set{N}_\op{c}$, the number of possible $\set{N}_\op{l}$ is finite. When $n$ is selected randomly and Algorithm~\ref{alg:general_alg} does not converge, there must be one node $n_0$ with $\Delta u(n_0,\set{N}_\op{l}) > 0$ which never gets selected. The probability of this happening goes to zero as the number of iterations goes to infinity.
\end{proof}

\begin{figure}
  \begin{center}
\def\svgwidth{0.6\columnwidth}
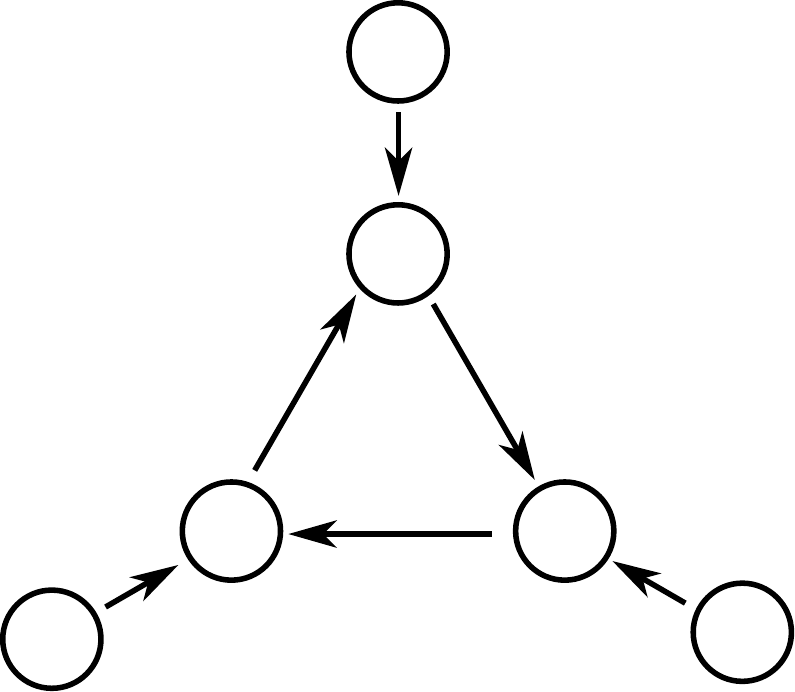
 \caption{Graph for which Algorithm~\ref{alg:general_alg} with $\Delta u$ as in \eqref{eq:delta_u_g} does not converge. }
\label{fig:circular_graph_game_limit_cycle.pdf}
\end{center}
\end{figure}

Using the individual objective \eqref{eq:delta_u_g}, Algorithm~\ref{alg:general_alg} is not guaranteed to converge to an equilibrium. The following example illustrates a case.
\begin{example}\label{ex:limit_cycle}
 Figure~\ref{fig:circular_graph_game_limit_cycle.pdf} shows a graph $\set{G}_\op{c}$ for which Algorithm~\ref{alg:general_alg} does not converge with criterion \eqref{eq:delta_u_g}. When the algorithm starts, the gain from becoming a CL is positive and the same for nodes 1, 2, and 3. Assume that node 1 is selected. Then nodes 3 and 4 become its CFs. Assume $\rho_\op{f}/\rho_\op{l} > 1.1$. Then it is not beneficial for node 3 to become a CL. Since there is no CL for node 2, it is beneficial for node 2 to become one. Since $\rho_\op{f}/\rho_\op{l} > 1.1$, it is more beneficial for node 1 to become a CF. Then node 3 becomes a CL and so on. If $\rho_\op{f}/\rho_\op{l} < 1.1$, we get a similar cyclic behavior with reversed direction.
\end{example}

\section{Simulations}
\label{sec:simulations}

\begin{figure}
  \begin{center}
 \input{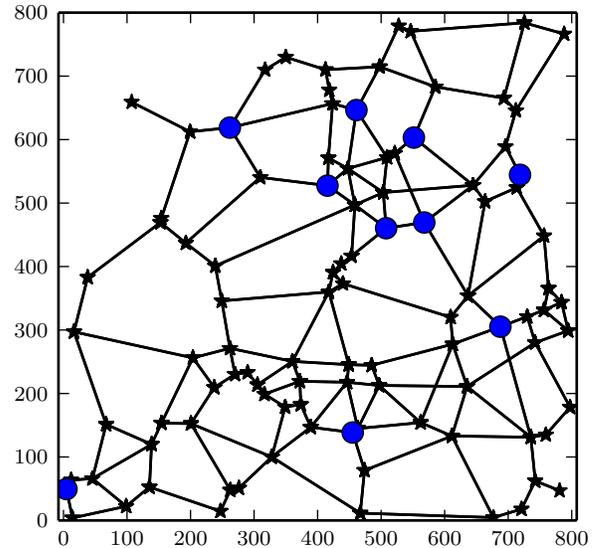}
 \caption{Randomly generated road network used in the simulation. Start/destination nodes are marked by a blue circle.}
\label{fig:road_network}
\end{center}
\end{figure}

In this section we present a simulation study of the platoon coordination method developed in this paper.
The road network that was used in the simulations was generated randomly by sampling 100 location points uniformly in a square of side length 800. All combinations of these locations were sorted by their Euclidean distance. Then, starting from the combination with the shortest distance, the combinations of locations were connected by two road segments (one in each direction) with length according to the Euclidean distance, if there had not already been a path between the two locations that was at most $1.5$ times longer than the Euclidean distance. The road network is shown in Figure~\ref{fig:road_network}.

We evaluated Algorithm~\ref{alg:general_alg} in four variants (greedy/random node selection and total/pairwise gain) with Monte Carlo simulations. In the figures and the following discussion we use ``total''/``pairwise'' to refer to $\Delta u$ according to \eqref{eq:delta_u_c}/\eqref{eq:delta_u_g}, and ``greedy''/``random'' to refer to selecting the node with largest $\Delta u$ or a random node in each iteration. 
We considered a nominal speed of $80$ according to which arrival times were set,  i.e., the constant speed of a CL $j$ is $v_j = 80$.  
For the fuel model we considered that $F^0 = 1$, $F^1 = 1/80$, and $F_p^0 = .9F^0$, $F_p^1 = .9F^1$. We get that \eqref{eq:v_star_def} evaluates to $80 \pm 35.777 = (44.223, 115.777)$.
For pairwise gain we terminated Algorithm~\ref{alg:general_alg} additionally once a specific $\set{N}_\op{l}$ reoccurred in order to avoid ending up in an infinite loop as described in Example~\ref{ex:limit_cycle}. 

\begin{figure}[t]
  \begin{center}
 \input{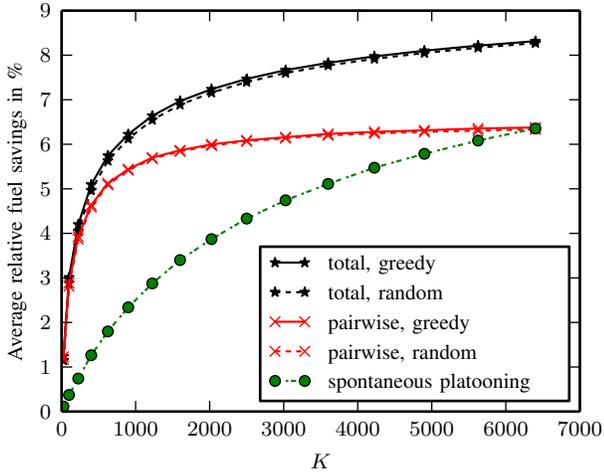}
 \caption{Average relative fuel savings for different numbers of trucks $K$. }
\label{fig:fuel_savings_K}
\end{center}
\end{figure}

\begin{figure}[t]
  \begin{center}
 \input{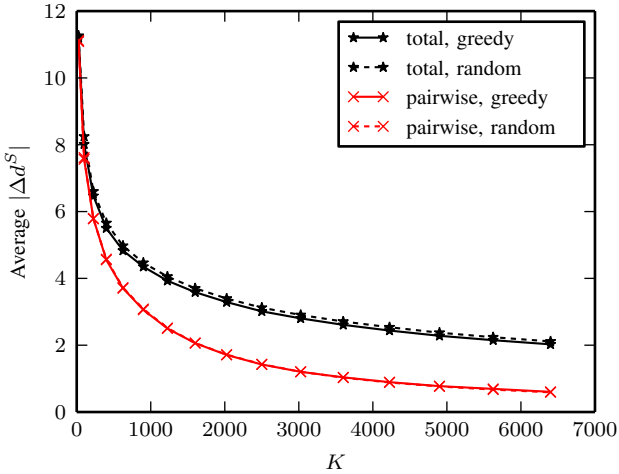}
 \caption{Average $|\Delta d^\op{S}|$ for different numbers of trucks $K$. }
\label{fig:delta_d_K}
\end{center}
\end{figure}

\begin{figure}[t]
  \begin{center}
 \input{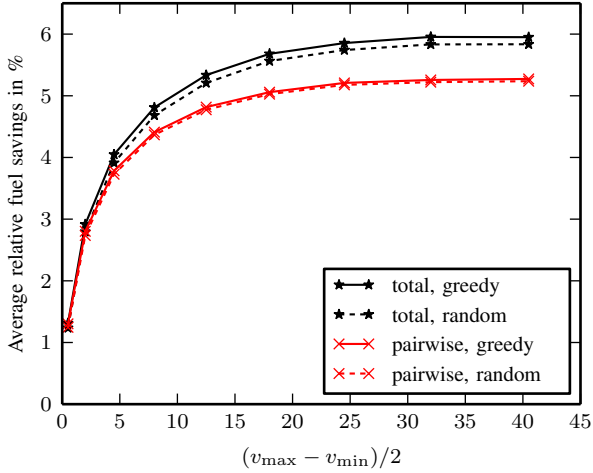}
 \caption{Average relative fuel savings for different $v_{\max} - v_{\min}$ with $K = 400$. }
\label{fig:fuel_savings_delta}
\end{center}
\end{figure}

We conducted simulations for different numbers of trucks $K$. For each $K$, 100 simulations were conducted. The starting times were sampled uniformly in the interval of $[0,1]$. Start and destination nodes were randomly selected from a subset of 10 nodes. The arrival times were calculated assuming a speed of 80. We set $v_{\min} = 70, v_{\max} = 90$. 
In each simulation we ran Algorithm~\ref{alg:general_alg} for the four combinations of using either total or pairwise gain, and either greedy or random node selection. For the pairwise gain we used $\rho_\op{l} = \rho_\op{f} = 0.5$. 
Figure~\ref{fig:fuel_savings_K} shows a plot of the fuel savings compared to the case where all trucks travel at a speed of 80 and do not platoon. Additionally we calculated fuel savings which would result from spontaneous platooning according to the following scheme. We assumed that all trucks travel with the nominal speed of 80. For each segment in the road network and for all trucks traversing a particular road segment, we collected the times when the trucks traverse the segment. Then, in the order of these points in time, we grouped trucks into platoons in such a manner that the difference in the traversal time within the platoon is at most $0.01$. We assumed that these platoons can platoon over the whole road segment with no coordination phase. Figure~\ref{fig:delta_d_K} shows $|\Delta d^\op{S}|$ averaged over all CL-CF pairs within one simulation and over the simulations for a specific $K$. 
Furthermore, we conducted simulations for different sizes of the band $[v_{\min}, v_{\max}]$ around the nominal speed of $80$ in which trucks can select their rendezvous speed. $K$ was fixed to $400$. Figure~\ref{fig:fuel_savings_delta} shows the relative fuel savings for different sizes of the band averaged over 100 simulations for each size. 

\begin{figure}[t]
  \begin{center}
 \input{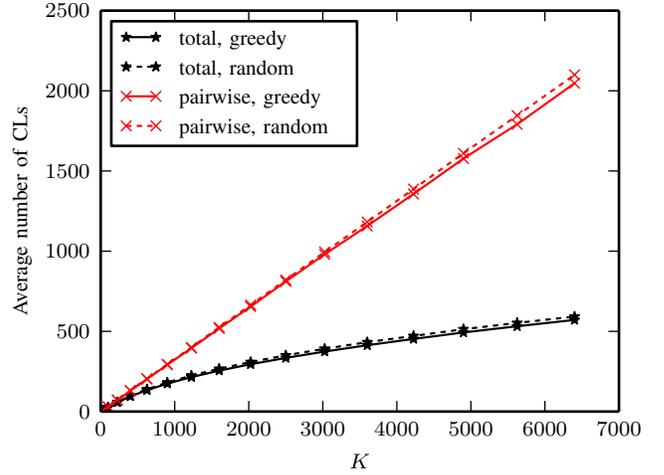}
 \caption{Average number of CLs for different numbers of trucks $K$. }
\label{fig:num_leaders_K}
\end{center}
\end{figure}

\begin{figure}[t]
  \begin{center}
 \input{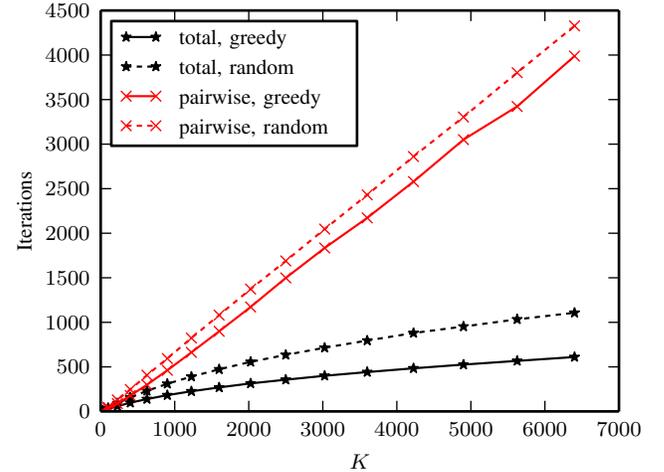}
 \caption{Average number of iterations until the algorithm terminates for different numbers of trucks $K$. }
\label{fig:iterations_K}
\end{center}
\end{figure}

In Figure~\ref{fig:fuel_savings_K} we can see that the relative fuel savings increase with the number of trucks. The relative fuel savings increase quickly with $K$ for small values of $K$ and then increase slowly for larger values. While the relative fuel savings with total gain keep increasing for large values of $K$, they are almost constant for pairwise gain. Note that the relative fuel savings in the simulations are upper bounded by $10 \%$. To reach this upper bound each truck needs to be a platoon follower during its entire journey. The difference between greedy and random node selection is small. For small $K$ the ratio between the relative fuel savings due to the coordination algorithm and those due to spontaneous platooning is relatively large and gets smaller for increasing $K$. Figure~\ref{fig:delta_d_K} shows that the average $|\Delta d^\op{S}|$ is relatively large for small $K$ and that it drops quickly to smaller values as $K$ increases. The values for pairwise gain are consistently smaller than those for total gain. 
Figure~\ref{fig:num_leaders_K} shows the average number of CLs for different $K$. We see that the number increases with $K$. For total gain it increases sub-linearly. The number of CLs is smaller for total gain compared to pairwise gain. For pairwise gain the number of CLs is proportional to $K$. The number of iterations until the algorithm terminates, shown in Figure~\ref{fig:iterations_K}, is almost proportional to the number of CLs, the difference being that random node selection leads to a significantly higher number of iterations than greedy node selection, in particular for total gain. 
Figure~\ref{fig:fuel_savings_delta} shows that fuel savings quickly increase for small $v_{\max} -v_{\min}$ and only increase moderately for larger bands.

We can conclude that the coordination of platooning is crucial for a small number of trucks and can significantly improve the overall fuel savings for a large number of trucks compared to spontaneous platooning. With coordination, already a small number of platooning enabled trucks can achieve significant reduction in fuel consumption. Pairwise gain creates more CLs and less coordination effort (smaller $|\Delta d^\op{S}|$) than total gain. Total gain yields superior performance. Random and greedy node selection give similar results but random node selection leads to more iterations. However, random node selection might be interesting for a distributed and parallel implementation since it does not require coordination amongst all nodes to determine which node is updated. 
Relatively small adjustments of the speed are sufficient to achieve most of the fuel savings possible. One should also keep in mind that fuel consumption per distance traveled is highly non-linear over the whole range of speeds a truck can attain and a first order fuel model is only accurate in a small range of speeds. 
Fuel consumption per distance traveled changes with speed in such a way that we expect large adjustments of the speed to be even less beneficial than predicted by the linear model.

\section{Conclusions and Future Work}

In this paper we started from a hybrid model of a single truck traveling in a road network. Using a first order fuel model, we derived a pairwise plan for two vehicles. Those pairwise plans, evaluated for all pairs of vehicles, served as the input data to a clustering algorithm that determines a set of coordination leaders. All other trucks in a cluster adapt their speed profile to the coordination leader of that cluster. We proposed four different variants of this algorithm and analyzed their convergence properties. We showed in simulation that this method can quickly find plans for a large number of trucks. The simulations also gave insight in how our approach behaves for different number of trucks in the network and different velocity constraints and supported that coordination of platooning is crucial to leverage its full potential to save fuel.

In future work we are interested in characterizing the equilibria of the algorithm analytically and establishing bounds on the performance. We are interested in studying the algorithm when applied in a receding horizon fashion where the plans are recalculated repeatedly. We plan to consider more complex pairwise plans that feature more complicated dynamics, different speed limits, traffic, etc. Finally, we plan to test the algorithm on real data and look into distributed implementations.

\appendix 

\paragraph*{Proof of Proposition \ref{prop:unconstrained_vs}}

Let $d_\op{s} = x_1(t^\op{M}) - x_1(t^\op{S})$. Let $D = x_1(t^\op{Sp}) - x_1(t^\op{S})$. 
We have the relation
\begin{equation}
  d_\op{s} = \frac{v_\op{s}}{v_\op{s} - v_0} \Delta d. \label{eq:d_s_delta_d_relation}
\end{equation}
At time $t^\op{M}$ we have $x_0 = x_1$. After the meeting point, both trucks platoon at speed $v_0$. Assume that $1$ is the platoon leader. Hence the total fuel consumption of $0$ up to some distance from the current position $D$ which fulfills $D > d_\op{s}$ becomes
$
 f_0(v_\op{s})d_\op{s} + f_p(v_0)(D - d_\op{s})
 = (f_0(v_\op{s}) - f_p(v_0))d_\op{s} + f_p(v_0)D.
$
The fuel consumption of $1$ is not affected by $v_\op{s}$. We see that the term $f_p(v_0)D$ is not a function of $v_\op{s}$, so the optimal rendezvous speed does not depend on the total distance traveled. In order to find the optimal $v_\op{s}$, we can therefore consider the remaining terms denoted as $f_r(v_\op{s})$ and get with \eqref{eq:d_s_delta_d_relation} and the definitions of $f_0$, $f_p$
\begin{align*}
 f_r(v_\op{s}) &= (f_0(v_\op{s}) - f_p(v_0))d_\op{s}\\
 &= (F^1 v_\op{s} - F_p^1 v_0 + \Delta F^0) \frac{v_\op{s}}{v_\op{s} - v_0}\Delta d,
\end{align*}
with $\Delta F^0 = F^0 - F_p^0$. We take the derivative of the above expression in order to find its extrema
\begin{align*}
 &\frac{\partial}{\partial v_\op{s}} f_r(v_\op{s}) = \\
 &= \frac{\Delta d}{(v_\op{s} - v_0)^2} (F^1 v_\op{s}^2 - 2F^1 v_0 v_\op{s} + F_p^1 v_0^2- \Delta F^0 v_0).
\end{align*}
In order to find the extrema $\tilde{v}_\op{s}$, we check where this expression is zero. We can assume that $\Delta d \neq 0$, otherwise $d_\op{s} = 0$ which means that the trucks can directly start platooning. Therefore 
\begin{align}
 0 &= (F^1 (\tilde{v}_\op{s})^2 - 2F^1 v_0 \tilde{v}_\op{s} + F_p^1 v_0^2- \Delta F^0 v_0) \label{eq:optimal_vs_cond}\\
 \tilde{v}_\op{s} &= v_0\left(1 \pm \sqrt{1 - \frac{F_p^1}{F^1} + \frac{\Delta F^0}{F^1 v_0}}\right).\label{eq:v_star_def}
\end{align}
We have to differentiate between two cases. Either $\Delta d > 0$ which implies $v_\op{s} > v_0$, i.e., the CF speeds up, or $\Delta d < 0$ which implies $v_\op{s} < v_0$, i.e., the CF slows down. Otherwise $d_\op{s}$ becomes negative. There are two solutions for $\tilde{v}_\op{s}$, one where $\tilde{v}_\op{s} > v_0$, and the other $\tilde{v}_\op{s} < v_0$. The appropriate one depending on $\Delta d$ is $\tilde{v}_\op{s}$, the optimal unconstrained rendezvous speed.

We can verify that this is indeed a minimum by considering the asymptotic behavior of $f_r(v_\op{s})$ when $f_r(v_\op{s})$ approaches $\pm \infty$ and when it approaches $v_0$. Assume $\Delta d > 0$ so that $\tilde{v}_\op{s} > v_0$. We have
$
 \lim\limits_{v_\op{s} \rightarrow \infty} f_r(v_\op{s}) = \infty$,$
 \lim\limits_{v_\op{s} \rightarrow v_0^+} f_r(v_\op{s}) = \infty
$
where we used that $f_0(v_0) > f_p(v_0)$ so that the term $f_0(v_0) - f_p(v_0)$ becomes positive, which is the prerequisite to save fuel by platooning. 
When we have $\Delta d < 0$, so that $\tilde{v}_\op{s} < v_0$, then
$
 \lim\limits_{v_\op{s} \rightarrow -\infty} f_r(v_\op{s}) = \infty$,$
 \lim\limits_{v_\op{s} \rightarrow v_0^-} f_r(v_\op{s}) = \infty.
$
This shows that if $\tilde{v}_\op{s} > v_{\max}$, then $v_\op{s}^* = v_{\max}$, if $\tilde{v}_\op{s} < v_{min}$, then $v_\op{s}^* = v_{\max}$, and $v_\op{s}^* = \tilde{v}_\op{s}$ otherwise.

In order to have real solutions for \eqref{eq:optimal_vs_cond}, we need
\begin{align*}
 1 - \frac{F_p^1}{F^1} + \frac{\Delta F^0}{F^1 v_0} &> 0 \Leftrightarrow
 F_p^1 v_0 + F_p^0 < F^1 v_0 + F^0\\
 \Leftrightarrow f_p(v_0) &< f_0(v_0),
\end{align*}
which is the condition that the CF saves fuel when platooning. The larger the difference $f_0(v_0) - f_p(v_0)$ the larger the absolute difference between $v_0$ and $v^*$, i.e., the longer the trucks platoon. 

\bibliography{citations}
\bibliographystyle{IEEEtran}

\end{document}